# Rheology primer for nanoparticle scientists

Luigi Gentile* & Samiul Amin†

*University of Bari "Aldo Moro" and CSGI, Department of Chemistry, Via Orabona 4, 70126, Bari, Italy – luigi.gentile@uniba.it

†Manhattan College, Chemical Engineering Departmen, 3825 Corlear Avenue -Leo 423 Riverdale, NY 10463, USA - samin01@manhattan.edu

## Contents

## Notation

$\gamma \rightarrow strain$

$\sigma \rightarrow stress$

$\dot{\gamma} \rightarrow shear\ rate$

$\omega \to angular\ frequency$

$\tau \to observation\ time$

$\tau_M \to$ characteristic relaxation time of Maxwell

$G'(\omega) = G' \to storage\ or$ elastic modulus

$G''(\omega) = G'' \to loss\ or$ viscous modulus

$\xi \to$ correlation lenght

$\eta^*(\omega) = \eta^* \to complex\ viscosity$

$\eta(\dot{\gamma}) = \eta \to shear\ or\ dynamic\ viscosity$

$|\eta^*(\omega)|_{\omega \to 0} \to zero-frequency\ viscosity$

$|\eta(\dot{\gamma})|_{\dot{\gamma} \to 0} = \eta_0 \to zero-shear\ viscosity$

$|\eta(\dot{\gamma})|_{\dot{\gamma} \to \infty} = \eta_\infty \to infinity-shear\ viscosity$

$\lambda \to characteristic\ time\ of\ the\ Carreau-Yasuda\ model$

$De \to Deborah\ number$

$Pe \to Peclet\ number$

$Re \to Reynold\ number$

$\Delta r(t) \to mean\ squared\ dispacement$

$g_1(\tau) \to electric\ field\ autocorrelation\ function$

## 11.1 Principles of Rheology

The art of observation is the active acquisition of information from a primary source. A comprehensive approach to observation, inquiry, and measurement is *perspectival observation*, a term adapted from Schwartz and Ogilvy (1979),[1] which connotes not anything-goes-subjectivity, but the differences in perspective, which arise when observing from a different "place" - "place" meaning everything from geographical position to psychologically-embedded history. Perspectival observation also connotes that the measuring process may alter the phenomenon being measured.[2] Mechanical properties are investigated by applying an oscillatory or stationary deformation. In the first, case the measurements in made to minimizing the effect of the measuring process on the material, while in the latter the measured property (viscosity) is a function of the deformation. The reader will soon appreciate that mechanical properties can be related to the structural parameters of the colloidal suspensions such as particles shape and size. Moreover, the mechanical properties are very sensitive to transitions for instance fractal aggregation of colloidal particles and consequently percolation. Several products contain a certain fraction of particles such as inks, cosmetic, pharmacological and personal care products or even biological fluids as a consequence understanding of the viscoelastic behaviour of these suspensions is relevant to improve properties and production properties.

The simplest geometry generally adopted to introduce general terms is the two-plate model. In which the material under investigation is in between the two plates placed at distance $h$. The bottom plate is stationary while the upper plate can be moved parallel to the bottom plate, Figure 11.1. The *shear stress*, $\sigma$, is defined as the ratio between the applied force, $F$, and the surface of the upper plate, $A$ on which the force is applied. A linear elastic solid follows the Hooke law, i.e. $\sigma = G(\Delta l/h)$ where $G$ is the characteristic elastic constant of the solid and $\Delta l$ is the plate displacement, the $\Delta l/h$ ratio is the so-called *strain*, $\gamma$. The distance between the plates, $h$, has to be small enough in a way that the strain is constant within the sample thickness, i.e. laminar flow condition. A linear viscous fluid follows the Newton law, i.e. $\sigma = \eta(d\gamma/dt)$ where $\eta$ is the characteristic viscosity of the fluid and $d\gamma/dt$ is the so-called *shear rate*, $\dot{\gamma}$. The shear rate is defined as the velocity gradient between two moving surfaces $\dot{\gamma} = dV/dh$ that for laminar flow will be $\dot{\gamma} = V/h$.

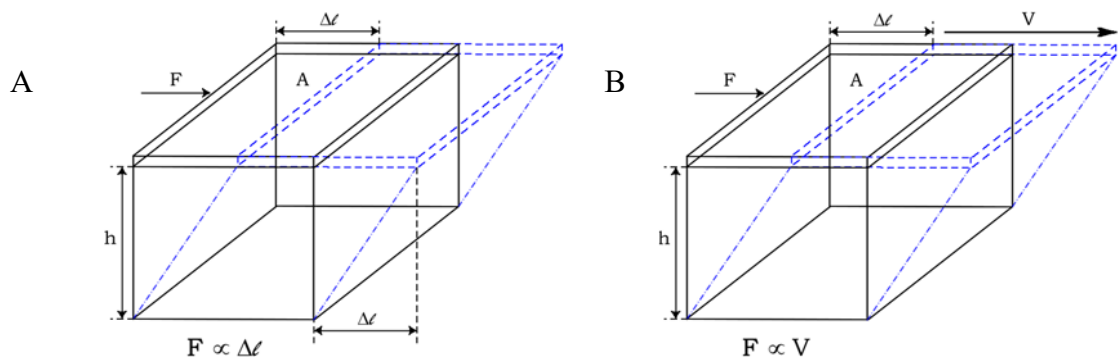

**FIGURE 11.1** Two-plate model: The sample is placed between two plates, the bottom plate is stationary while the upper plate can be moved parallel to the bottom plate. (A) The applied force, $F$, is proportional to the plate displacement, $\Delta l$, for a linear elastic solid; (B) The applied force is proportional to the velocity of deformation, $V$, for a linear viscous liquid.

However, it is essential to consider the characteristic relaxation time, $\Lambda$, of each material respect to the applied deformation (or stress) and the observation time, $\tau$, for which such deformation is applied. The ratio between the relaxation time, $\Lambda$, and the observation time, $\tau$, is the so-called Deborah number, $De = \Lambda/\tau$. Introduced by Markus Reiner (1964) inspired by a verse in the Bible, stating, "*The mountains flowed before the Lord*". $\Lambda$ is the characteristic time of a material needed to respond the applied deformation, at lower Deborah numbers, the material behaves in a more fluid-like manner, with an associated Newtonian viscous flow. At higher Deborah numbers, the material behavior enters the non-Newtonian regime, increasingly dominated by elasticity and demonstrating solid-like behavior.[3] This approach incorporates both the elasticity and viscosity of the material, i.e. its rheological behaviour. Rheology is the science of flow and deformation of matter and describes the interrelation between force, deformation and time. Eugene C. Bingham (1920) introduced the term rheology following a suggestion of Markus Reiner. A solid can be defined as $De \rightarrow \infty$, while a liquid $De \rightarrow 0$, in between the viscoelastic materials. To investigate viscoelastic materials dynamic or oscillatory measurements are needed. There is a wide range of users addressing rheology as the viscosity of the material, such a limited vision of the technique leads to common misinterpretation of the viscoelastic properties. The following will address first the dynamic or oscillatory measurements and subsequently the stationary measurements.

## 11.2 Oscillatory measurements

To know the intrinsic mechanical nature of a material a *small-amplitude oscillatory shear* (SAOS) should be applied to the material. In other words oscillatory shear stress (a force per unit area), $\sigma$, or a strain (deformation in a space), $\gamma$, with a certain angular frequency (in rad/s), $\omega$, is applied on the axis of, for instance, a cone of the conventional cone-and-plate geometry per a time, $t$, eq. 11.1,

$$\gamma(t) = \gamma_0 \sin(\omega t) \tag{11.1}$$

where $\gamma_0$ is the strain amplitude that for a SAOS experiment will be typically below 1%. The resulting shear stress produced by a small-amplitude deformation is proportional to $\gamma_0$ and sinusoidal varying in time, for an ideal linear elastic solid will be

$$\sigma(t) = G\gamma_0 \sin(\omega t) \tag{11.2}$$

Where $G$ is known as the *spring constant* or elastic modulus of the ideal solid defined by the Hooke law, $\sigma = G\gamma$. On the other hand, an ideal linear viscous liquid will answer to the strain deformation (eq.11.1) with the following

$$\sigma(t) = \eta\gamma_0\omega \cos(\omega t) \tag{11.3}$$

where $\eta$ is the viscosity that for an ideal linear viscous liquid is the equivalent of the one defined by the Newton law, $\sigma = \eta\dot{\gamma}$, where $\dot{\gamma}$ is the shear rate, i.e. $d\gamma/dt$.

However, the majority of the materials has the following stress response to the oscillatory strain deformation of eq.11.1

$$\sigma(t) = \sigma_0 \sin(\omega t + \delta) \tag{11.4}$$

where $\delta$ is the phase angle between the sinusoidal strain and stress. Eq.11.4 can be easily rewritten

$$\sigma(t) = \gamma_0[G'(\omega)\sin(\omega t) + G''(\omega)\cos(\omega t)] \tag{11.5}$$

where $G'(\omega) = (\sigma_0/\gamma_0)\cos(\delta)$ is the storage modulus (so-called elastic modulus) and $G''(\omega) = (\sigma_0/\gamma_0)\sin(\delta)$ is the loss modulus (so-called viscous modulus). The unit for $G'$, $G''$ and shear stress is 1 N/m² = 1 Pa (pascal) in the international system of units (SI).

Two kind of experiments can be performed by fixing the angular frequency $\omega$ and changing the amplitude $\gamma_0$ as a function of time (amplitude sweep experiment) or fixing the amplitude and changing the angular frequency (frequency sweep experiment).

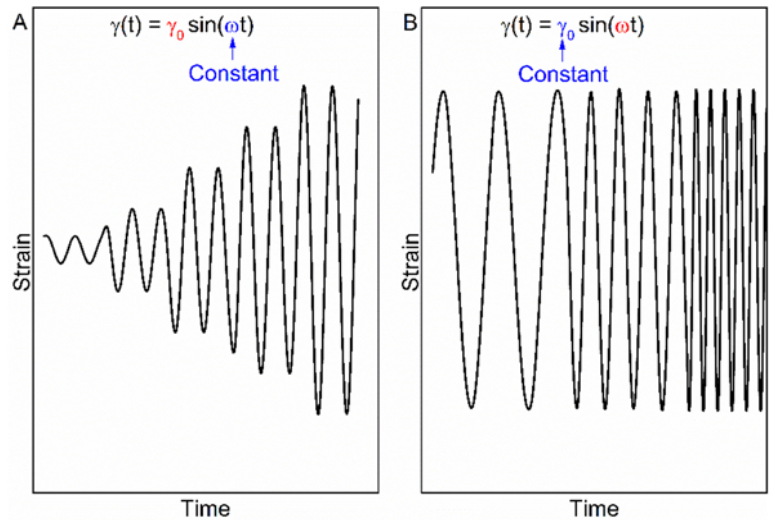

**FIGURE 11.2** Present of an amplitude sweep (A) and a frequency sweep (B). In the amplitude sweep the strain increases in five steps while the angular frequency is kept constant at all five measuring points. In the frequency sweep the frequency increase in three steps while the strain amplitude is kept constant.

### 11.2.1 Amplitude sweep experiment

The **strain- or stress-amplitude sweep experiment** is obtained by fixing the angular frequency and changing the applied initial strain or stress amplitude. This experiment is extremely useful to determine [4,5]:

(i)  The linear viscoelastic regime in which the test can be carried out allowing the sample to restore the deformation (it is function of the characteristic relaxation time of the material);
(ii) The state of the material, i.e. gel-like or liquid-like;
(iii) The apparent yield stress;
(iv) Shear-induced structures or anyway upper boundary before applying *large-amplitude oscillatory shear* (LAOS).

Figure 11.3 shows two most common strain amplitude sweep behaviour for viscoelastic materials; in both of them the linear viscoelastic regime (LVE) is highlighted by a dark grey zone on which $G'$ and $G''$ are independent respect to the applied strain, they are only function of the applied angular frequency $\omega$, i.e. in the *small amplitude oscillatory shear* (SAOS) where $G' > G''$. $G'$ and $G''$ after the SAOS regime are function of both $\omega$ and $\gamma$, in the *medium and large amplitude oscillatory shear* (MAOS and LAOS). In the MAOS regime figure 11.3A reports a broad peak for $G''$ just before the cross over between $G'$ and $G''$ that almost matches with the maximum in the shear stress profile indicating the presence of a yield point, $\gamma_y$. The yield point is the stress beyond which a material becomes plastic. The values of the viscous modulus $G''$ describe the portion of the deformation energy that is lost by internal friction during deformation. The elastic modulus $G'$ initially dominates over $G''$ until the cross over, consequently the $G''$ dominates and the entire material will start to flow, often named the point of complete fluidization, $\gamma_f$ [6]. Subsequently in the LAOS regime the plastic behaviour grow further, the system is fully fluidized, and the distance between the moduli grows. Figure 11.3B shows a no yielding material, in fact there is no shear stress peak before the crossover ($G' = G''$), while a peak is observed in the LAOS regime together with a peak on both elastic and viscous moduli, such behaviour can be associated with the formation of new structures induced by the deformation. For instance, the formation of multilamellar vesicles (MLVs) from planar lamellae.[7]

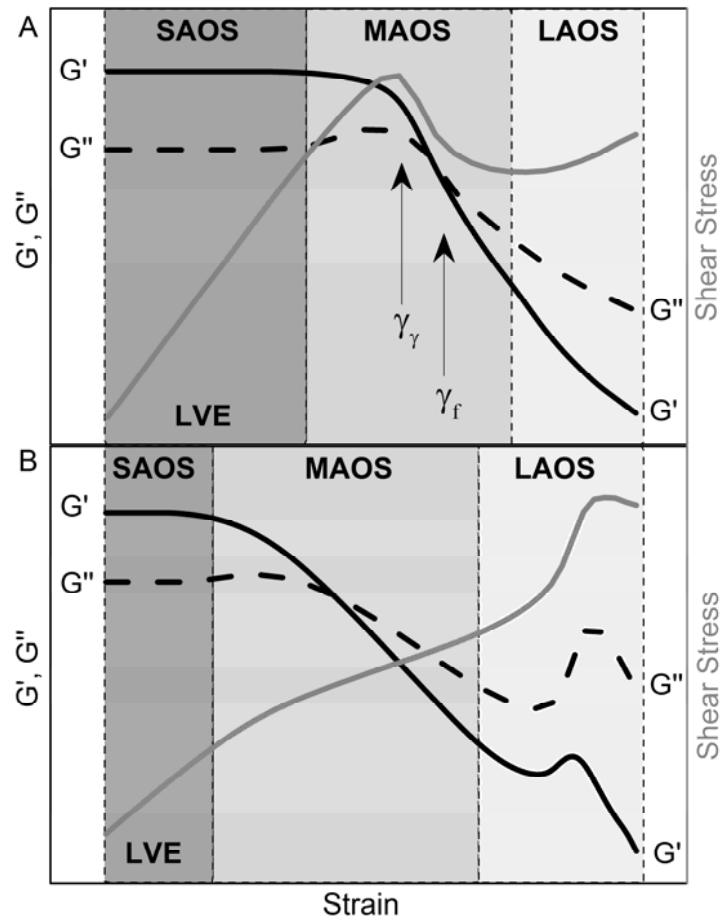

**FIGURE 11.3** Strain amplitude sweep experiments of two generic systems, a yielding material (A) and a non-yielding material subject to deformation-induced transitions (B). All axis are in the logarithmic scale, $G'$ is depicted by the black solid line, while $G''$ by black dash line and shear stress by solid grey line. The yield strain point is indicated by $\gamma_y$, while the point of complete fluidization is indicated by $\gamma_f$.

Emulsions exhibit a broad range of different rheological properties, ranging from low viscosity liquids to viscoelastic solids. The terms colloid and emulsion are sometimes used interchangeably, however emulsion should be used when both phases, dispersed and continuous, are liquids, i.e. emulsions are suspensions of droplets of one immiscible fluid in another. The droplets are typically stabilized by a surfactant adsorbed on their interfaces (micro-emulsions); this ensures that there is a short-range repulsive interaction between them, which prevents coalescence. The amplitude sweep of repulsive emulsions above the random close packing, (the volume fraction, $\phi$, for a random close packing is approximately 0.64 [8,9]) exhibits yield stress and $G''$ exhibits a single well-defined peak at the same strain.[10] On the other hand, attractive emulsions, i.e. characterized by droplets with additional attractive interactions, begging to yield at strain values lower than the respective repulsive ones (below ~1%) and $G''$ exhibits two well-defined peaks before falling with a slope that is ~half of the $G'$ slope.[10]

### 11.2.2 Frequency sweep experiment

The **frequency sweep experiment** is performed over a range of oscillation frequencies at a constant oscillation (strain or stress) amplitude in the linear viscoelastic regime (LVE), i.e. SAOS. This is the most adopted experiment to characterize materials and their properties without destroying the microstructure. Figure 11.4 shows an example of a full frequency sweep behaviour, even though often only one or two regions are detected by the classical rheological experiments. To visualize a full frequency sweep micro-rheology can be adopted (see paragraph 11.4.3) or a master curve can be constructed by using experiments made for instance at different temperatures.[5]

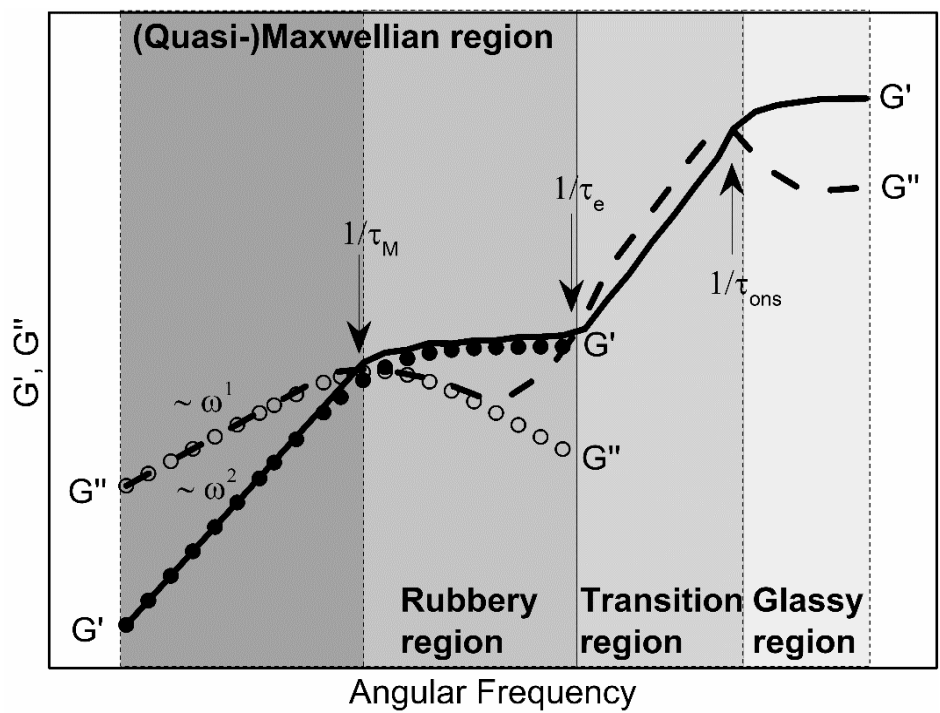

**FIGURE 11.4** Frequency sweep experiment of a generic system performed in the LVE region. The $G'$ (solid black line) and $G''$ (dashed black line) are plotted against the angular frequency in a log-log diagram. Four regions can be identified as well as 3 cross-over between the elastic and viscous moduli. The first two regions can be described by a quasi-Maxwellian behaviour; $G'$ (filled circles) and $G''$ (open circle) of the Maxwell model are shown in figure.

The frequency sweep depicted in Figure 11.4 can be divided in 4 regions.

(I) The first region shows the typical "liquid-like" fluid behaviour where $G' \sim \omega^2$ is much lower than the $G'' \sim \omega^1$, sometimes this region is called *terminal region*. Here a plateau in the complex viscosity $|\eta^*| = \sqrt{(G''/\omega)^2 + (G'/\omega)^2}$ can be observed.

Assuming that the Cox-Merz rule, i.e. $|\eta^*(\omega)|_{\omega \to 0} = |\eta(\dot{\gamma})|_{\dot{\gamma} \to 0}$ [9] is valid than the *zero-shear viscosity* can be measured.

(II) The second region also called "rubber elastic plateau" shows a typical viscoelastic "gel" behaviour in which $G'$ is almost constant respect to ω and always higher than $G''$. For polymers this region will be one on which their entanglements are beginning to form a temporary network. The cross over between the moduli $G' = G''$ is the characteristic relaxation time.

The Maxwell model can very well describe the first and second regions except for the deviation at high angular frequency. The simple Maxwell model considers both the $G'$ and $G''$ moduli of the material by assuming the rheological properties as represented by a spring (elastic component) and a dashpot (viscous component) connected in series [4,5,11] leading to

$$G'(\omega) = G_0 \frac{\omega^2 \tau_M^2}{1 + \omega^2 \tau_M^2}, \qquad G'(\omega) = G_0 \frac{\omega \tau_M}{1 + \omega^2 \tau_M^2} \qquad (11.6)$$

where $G_0$ is the elastic modulus at its high-frequency plateau, while the characteristic relaxation time of Maxwell is $\tau_M$ (first crossover in figure 11.4). Figure 11.4 is a generic representation of a typical frequency sweep, however, it should be pointed out that there are different behaviour, for instance, a polymer described by the Zimm model [12] will not show any $\tau_M$ which implies no crossover between $G'$ and $G''$. Moreover, knowing $G_0$ and $\tau_M$ the *zero-frequency viscosity* can be determined or reversely knowing $\tau_M$ and $|\eta^*(\omega)|_{\omega \to 0}$, $G_0$ can be determined [13–16]

$$|\eta^*(\omega)|_{\omega \to 0} = G_0 \tau_M \qquad (11.7)$$

$G_0$ can be particularly relevant in surfactant science, since can be related to the correlation length ξ (i.e. the mesh size of a bicontinuos phase) [13–17]

$$\xi \approx \sqrt[3]{\left(\frac{k_B T}{G_0}\right)} \qquad (11.8)$$

In polymer science the $G'$-$G''$ crossover is usually related to the molecular weight of the polymer.[5] More interesting by following the Doi and Edwards [18] description of the de Gennes [19] reptation theory one can say that at the $\tau_M$ the many-body system of a melted or concentrated solution of polymer is simplified to a single-chain problem, i.e. at that is the time at which the center of mass of the chain diffuse a distance on the order of the chain's own coil size $R$, the chain would need to move a distance called primitive path, $L$, along its contour in the tube (average contour length of the tube) of the Doi and Edwards tube model, eq.11.9

$$\tau_M \sim \frac{L^2}{D_t} \qquad (11.9)$$

Where $D_t$ is the (three-dimensional) self-diffusion coefficient. Eq. 11.9 can be adapted also for rigid rods in which $D_R$ will be the parallel diffusion coefficient and $L$ the length of the rod.[6]

(III) The transition region is characterized by slightly higher values of $G''$ respect to $G'$ and their growth as a function of the angular frequency at a similar rate, sometimes called flow zone or Rouse region. This region is preceded by a $G''$-$G'$ crossover, the reciprocal is historically labeled as the tube confinement time or the entanglement time $\tau_e$ which defines the onset of entanglement effects in polymers melt or solutions [20,21]

$$\tau_M = \tau_e \left(\frac{N}{N_e}\right)^3 = \tau_e \left(\frac{L}{l_{ent}}\right)^3 = \tau_e (Z)^3 \qquad (11.10)$$

where $N$ is the number of Kuhn elements and $N_e$ is the number of Kuhn segments between entanglement points, while $L$ is the contour length and $l_{ent}$ is the entanglement spacing therefore $Z$ is defined as the number of entanglement points.

(IV) In the high-frequency region a new $G'$-$G''$ crossover is observed before the moduli reach a steady-state value and the distance between them increases. This state is commonly addressed as the "glassy state" this definition is again coming from the rheology of polymers and reciprocal of the crossover in the frequency sweep is the time above which polymer-specific relaxations become relevant, $\tau_{ons}$.

The frequency sweep of repulsive emulsions can result in a viscoelastic behaviour above the random close packing ($\phi > 0.64$) but it can also result in a fluid-like behaviour when the effective volume fraction, $\phi_{eff}$, decreases below 0.64.[10] The effective volume fraction that focuses on the consequences of the droplet packing independently from the repulsive or attractive interactions is defined as $\phi_{eff} \approx \phi(1 + (3\delta/2R))$, where $\delta$ is the thickness of the surfactant layer adsorbed at the droplet surfaces and R the droplet radius.[22] On the other hand, attractive emulsions show always a viscoelastic behaviour in which $G'(\omega)$ is independent of $\omega$.[10] Mason and Weitz [22] proposed a model to describe $G'(\omega)$ and $G''(\omega)$ as a function of $\omega$ for which an analytical expression is provided by Mewis and Wagner.[23]

### 11.3 Stationary measurements

In the previous paragraph, a generic material was under oscillatory deformation, while a stationary deformation, i.e. rotational tests will be the focus of this paragraph. The rotational tests can be performed with two operation modes, viz. controlled shear rate or controlled shear stress. In both operational modes two tests can be performed: (i) fixing the shear rate or shear stress to monitor the viscosity or the stress as a function of time (step test); (ii) to monitor viscosity or shear stress as a function of the shear rate or the shear stress (flow curve).

#### 11.3.1 Flow curve

The most well-known rheological experiment is the so-called flow curve where the flow behaviour of a material is detected. Generally, the shear stress $\sigma(\dot\gamma)$ is monitored as a function of the shear rate, $\dot\gamma$, in a log-log scale. Most commonly, shear or dynamic viscosity, $\eta(\dot\gamma)$, is plotted against $\dot\gamma$ in isothermal conditions. The shear viscosity of a system is a measure of the resistance to flow of the material. A simple flow field can be established in a system by placing it between two plates and then pulling the plates apart in opposite directions at a rate that is the shear rate.

The most generic model to describe non-Newtonian fluid flow behaviour is the Herschel-Bulkley model [24]

$$\sigma(\dot\gamma) = \sigma_y + K(T)\dot\gamma^n \qquad (11.11)$$

where the consistency index at a fix temperature T is $K(T)$, $\sigma_y$ is the so-called yield stress point, while *n* is the flow index. Eq. 11.11 can also be written in terms of shear viscosity

$$\eta(\dot\gamma) = \sigma_y \dot\gamma^{-1} + K(T)\dot\gamma^{n-1} \qquad (11.12)$$

- when *n* < 1 and $\sigma_y \neq 0$ the fluid will be *shear-thinning* with yield stress or also called *yield pseudo-plastic*, i.e. the viscosity decreases non-monotonically increasing the shear rate when the shear stress overcome $\sigma_y$;
- when *n* < 1 and $\sigma_y \neq 0$ and $\sigma(\dot\gamma)$ overcome $\sigma_y$ the fluid is *shear-thinning* with a constant viscosity, $\eta_B$, consequently eq. 11.11 becomes $\sigma(\dot\gamma) = \sigma_y + \eta_B \dot\gamma$ than the fluid is called *Bingham plastic*; [25]
- when *n* < 1 and $\sigma_y = 0$ the fluid will be *shear-thinning* and consequently described with the simple power-law equation of Oswald and de Waele model,[26] this kind of fluid is also called *pseudo-plastic*;
- when n > 1 and $\sigma_y = 0$ the fluid will be *shear-thickening,* i.e. the viscosity increases increasing the shear rate such a behaviour is less common respect to the previous ones and it is usually due to in-flow collisions and Brownian motion or structural modification, for instance in lyotropic liquid crystals due to the relatively low bending rigidity of the bilayers.[27] This kind of fluid is called *dilatant*;
- when n = 1 and $\sigma_y = 0$ the fluid will be *Newtonian*, i.e. a fluid in which the shear stress is linearly correlated to the shear rate;
- when the viscosity or shear stress experience an unstable temporal behaviour, i.e. periodic oscillations [28] the Herschel-Bulkley model cannot describe this fluid behaviour due to flow instabilities.

However, the majority of the complex fluids show more than one of the behaviours discussed before in the observed shear rate region. It is often observed in the lowest shear rate region a *Newtonian-like* behaviour, *zero-shear* viscosity plateau, followed by a *shear-thinning* and ending with another *Newtonian-like* behaviour, *infinity-shear* viscosity plateau. In other cases, the zero-shear viscosity, $|\eta(\dot\gamma)|_{\dot\gamma \to 0}$ is not detected since out of the measurable range. These kind of behaviours are reported for polymer melts, polymer solutions, colloidal particles and rigid particles. The flow curve reported in Figure 11.5 are two generic examples that can be described by the Carreau-Yasuda model [5,29] in which for simplicity the *zero-shear* viscosity is indicated as $|\eta(\dot\gamma)|_{\dot\gamma \to 0} = \eta_0$ and the *infinity-shear* viscosity as $|\eta(\dot\gamma)|_{\dot\gamma \to \infty} = \eta_\infty$

$$\eta = \frac{\eta_0 - \eta_\infty}{(1 + (\lambda \dot{\gamma})^a)^{(1-n)/a}} + \eta_\infty \qquad (11.13)$$

where $a$ is the constant that determines the curvature of the transition region between the lower Newtonian regime (if $a = 2$ than Carreau model [29,30]), and the power law regime $n$ has the same meaning of the Herschel-Bulkley model [24], $\lambda$ has the dimension of a time representing a characteristic time of the system.

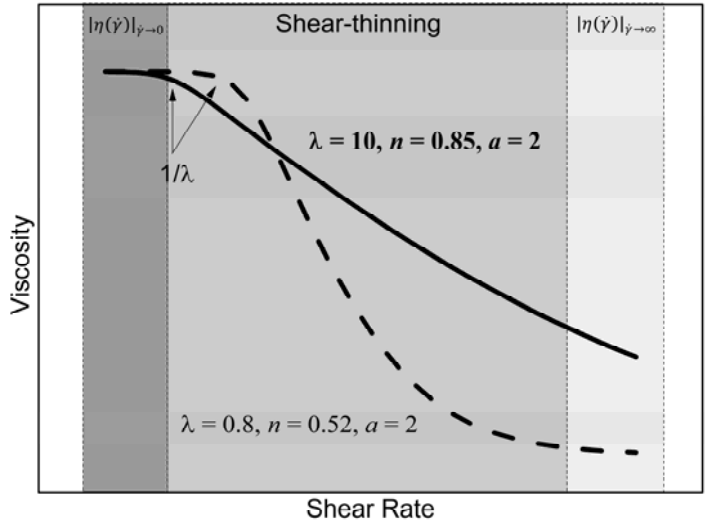

**FIGURE 11.5** Two generic flow curves in the log-log plot in which bold parameters describe the continuous line, while the others describe the dotted line. The *zero-shear* viscosity can be detected at low shear rates, while the *infinity-shear* viscosity at high shear rates. The $\lambda$ is the characteristic time in the Carreau-Yasuda model and $n$ is the exponent of the *shear-thinning* region.

### 11.3.2 The viscosity of particle suspensions

The simplest case consists in a suspension of monodisperse solid spheres with very low particle concentration. i.e. a volume fraction $\phi < 0.15$. The spheres are acting as obstacles in the fluid matrix increasing the dynamic viscosity. The *Newtonian-like* behaviours of the dynamic viscosity is described by the Einstein model,[31] eq. 11.14

$$\eta = \eta_s \left(1 + \frac{5}{2}\phi\right) \qquad (11.14)$$

where $\eta_s$ is the viscosity of the solvent. This model was modified by Batchelor [32] for higher volume fraction, in particularly for $0.15 \leq \phi \leq 0.2$, eq. 11.15

$$\eta = \eta_s \left(1 + \frac{5}{2}\phi + \frac{15.2}{2}\phi^2\right) \qquad (11.15)$$

The coefficient of the quadratic term 15.2/2 = 7.6 is the far-field hydrodynamic interaction. Several author has evaluated the coefficient of the quadratic term for instance 5.0 has been determined for random suspensions of hard-spheres in shear flow and the introduction of Brownian motion between particles increases the value to 6.0. [23]

One of the most successful model to describe suspensions with a higher volume fraction $\phi > 0.2$ is the Krieger & Dougherty model [33]

$$\eta = \eta_s \left(1 - \frac{\phi}{\phi_{max}}\right)^{-[\eta]\phi_{max}} \qquad (11.16)$$

In which $\phi_{max}$ is the upper limit for $\phi$, this limit is approximately 0.64 for random close packing and 0.74 for face-centered or hexagonal close-packed. $[\eta]$ in eq.11.16 is the intrinsic viscosity, such viscosity contains structural information of the system and in colloidal (particle) science is dimensionless. On the other hand, the *intrinsic viscosity* in polymer science is usually expressed in ml/g and it is defined as $[\eta] = |\eta_r(c)|_{c \to 0}$ where $\eta_r(c)$ is the *reduced viscosity* $\eta_r(c) = \eta_{sp}/c$ where $c$ is the concentration of the solute in g/ml and $\eta_{sp}$ the *specific viscosity* $\eta_{sp} = (\eta_0 - \eta_s)/\eta_s$ where $\eta_s$ is the viscosity of the solvent (dispersant) and $\eta_0$ is the *zero-shear* viscosity. Furthermore, the degree of dilution of a polymer solution can be expressed in terms of dimensionless Debye parameter, $c[\eta]$, in other terms the product of the intrinsic viscosity and the concentration of the solution is an approximation of the volume fraction $\phi$ of the polymer macromolecule in solution.

The *intrinsic viscosity* [η] for a suspension of particles is the factor that describes the particle shape, for a sphere 5/2 = 2.5, while for ellipsoids will be 2.94, 3.43, 7.00 for aspect ratio of 5:1, 10:1, 50:1 respectively. Eq. 11.14 can then be generalized respect to the particle shape

$$\eta = \eta_s(1 + [\eta]\phi) \qquad (11.17)$$

Eq. 11.14, 15, 16 and 17 are describing a *Newtonian-like* behaviour, however particle size, polydispersity and zeta potential (i.e. charge on the surface) along with the volume fraction are playing a role, in fact a *shear-thinning* behaviour can be observed for $0.1 \leq \phi \leq 0.5$ while high particle volume fraction ($\phi > 0.5$) can result in an yield stress.[34] At even higher volume fractions ($\phi \geq 0.6$), a *shear-thickening* can be detected due to cluster and jamming.[35,36] However, suspensions with small particles (< 1 μm) are particularly sensitive to changes in zeta potential [37] resulting in a higher dynamic viscosity. The size of the particles at the same volume fraction is mainly affecting the viscosity at low shear rate where small particles show higher viscosity due to the higher particle surface area (i.e. higher number of particles) that generates higher resistance to flow. Moreover, larger polydispersity maximise the maximum packing fraction, $\phi_{max}$, defined as the true volume of the particle divided by the apparent volume occupied by them when the packing is maximum, lowering the viscosity respect to the monodisperse case with the same volume fraction, $\phi$. In previous theoretical and experimental studies the viscosity behaviour of the suspensions was assumed to be dictated by the Peclet number (eq.11.18) and particle Reynolds number (eq.11.19), for instance Stickel & Powell (2005) [38] indicate *Newtonian* behaviour for values of $Pe \geq 10^3$ and $Re \leq 10^{-3}$.

$$Pe = \frac{6\pi\eta_s R^3 \dot{\gamma}}{kT} \qquad (11.18)$$

$$Re = \frac{\rho_s R^2 \dot{\gamma}}{\eta_s} \qquad (11.19)$$

where $\rho_s$ is the density of the solvent while $R$ is the radius of the particle. However, these numbers are related to the particles size, while polydispersity and zeta potential are not considered.

As discussed in paragraph 11.2.1 in case of liquid droplets (emulsions) the rheological behaviour is usually more variegate respect to solid particles. For high volume fraction $\phi > 0.45$ in the absence of appreciable colloidal interactions, the droplet size modifies emulsion rheology. Pal et al.[39] proposed the following equation for the *zero-shear* viscosity of the emulsion.

$$\eta_0 = \eta_s \left[ e^{\left(\frac{[\eta]\phi}{1-(\phi/\phi_{max})}\right)} \right]^K \qquad (11.20)$$

where $K = [(0.4 + M)/(1 + M)]$ and $M = \eta_d/\eta_s$ in which $\eta_d$ is the viscosity of the dispersed phase.

## 11.4 Microrheology:Introduction

The structure and dynamics in complex fluids and soft matter systems is complex and a effective quantitative characterization requires access to techniques which allow the structure and dynamics characterization over a wide range of lengthscales and timescales. Although as highlighted earlier, mechanical rheological techniques through the utilization of a stress or strain controlled rheometer allows insights into the mechanical properties, and dynamics of complex fluids, they are limited in the accessible frequency range. Mechanical rheometers cannot be reliably utilized to access short term dynamics as data at frequencies greater than 100Hz is not reliable due to inertia issues. Many of these complex fluids exhibit local heterogeneity in rheological behaviour, this local rheological behaviour cannot be accessed through a bulk measurement technique such as mechanical rheometery. Additionally many complex fluids, especially of biological origin such as proteins are only available in small quantities which makes utilization of mechanical rheometry challenging. These limitations in mechanical rheometry has led to rapid popularity of microrheological techniques.

Microrheology essentially involves the embedding of tracer particles in the complex fluid of interest to probe the rheological response.[40–49] This is accomplished through tracking the motion of the tracer particles in the complex fluid. The microrheology experiment can essentially still be envisioned to be very similar in concept as a mechanical rheometry experiment. In the case of microrheology the stress is the stress applied due to the motion of the particle, whilst the strain or deformation is the change in the tracer particles position.

### 11.4.1 Microrheology Classification: Active and Passive Microrheology

Microrheological techniques can be classified into two main categories based upon the type of stress experienced by the tracer or colloidal particle. These are:

- ➢ Passive Microrheology-in a passive microrheology measurement, no external stresses are exerted upon the tracer particle. The linear viscoelastic properties are extracted from the motion of the tracer particles undergoing thermal fluctuations. The data obtained is within the linear viscoelastic range and gives rise to elastic $G'$ and viscous $G''$ moduli over a wide frequency range covering several decades in frequency. This technique has been applied across a wide range of complex fluids materials classes such as polymers, proteins, surfactants.
- ➢ Active Microrheology-in an active microrheology measurement, an external stress is applied to cause the motion of the tracer particle. External fields such as Magnetic fields, laser tweezers (optical force) can be utilized to impart an external stress onto the tracer particle. Active microrheology allows the measurement to be not only carried out in the linear viscoelastic regime but can extend into the non-linear regime. In active microrheology measurements the tracer probe particle can be driven within the complex fluid either in the ocillatory or steady motion.

The following paragraphs will essentially focus on the passive microrheology techniques as those are the most widely used techniques and should be most relevant to researchers working in nanosciences.

### 11.4.2 Theoretical Background

The seminal work in the area was carried out by Mason and Weitz.[43,44] This was both from the perspective of the conceptual work as well as advancing experimental techniques to carry out microrheological measurements. The initial work carried out, connected the thermal fluctuations of the tracer probe particles to the frictional drag and therefore to the linear viscoelastic properties of the complex fluid in which the tracer probe particle is embedded in. The tracer probe particle, as quantified through the mean squared displacement MSD (Δr(t)), is connected to the rheological properties of the complex fluids through application of a Generalized Stokes Einstein Relationship (GSER) [43,44]

$$G^*(\omega) = \frac{k_B T}{\pi a <\Delta r\left(\frac{1}{\omega}\right)^2> \Gamma(1+\alpha(\omega))} \qquad (11.21)$$

where, $\Gamma$ is a gamma function due to Fourier transformation, ω is frequency, and α(ω) is a variable between 0 and 1. The storage modulus ($G'$) and the viscous modulus ($G''$) of the complex fluid can be obtained from the following expressions:

$$G' = G^*(\omega) \times \cos\left(\frac{\alpha(\omega)\pi}{2}\right) \qquad (11.22)$$

$$G'' = G^*(\omega) \times \sin\left(\frac{\alpha(\omega)\pi}{2}\right) \qquad (11.23)$$

A purely viscous material is indicated by the limit α(ω)=1 and $G'$=0, whilst a purely elastic material is indicated by α(ω)=0 and $G''$=0.

### 11.4.3 Experimental Techniques

The main objective of experimental techniques in any microrheology experiment is to obtain the MSD Δr(t) for the probe particles embedded in the complex fluid. This can be obtained through a variety of techniques, the main ones being light scattering techniques such as dynamic light scattering,[40] Diffusing Wave Spectroscopy [45,46] and microscopy based video particle tracking techniques.[47] These techniques have gained significant popularity and have been applied across a range complex fluids such as surfactant structured systems, [45,49] polymers,[46] clays [48] etc.

Light Scattering Techniques: Two light scattering techniques that have found extensive application in carrying out passive microrheology of complex fluids are dynamic light scattering (DLS) based microrheology and diffusing wave spectroscopy (DWS) based microrheology. Essentially both techniques rely on very similar principle of measuring the thermal motion of the tracer probe particles with coherent laser light. Probe particles moving around by Brownian motion scatter the light impinging upon them by a laser light source. The scattered light is captured by a photon detector from which an auto correlator gives the electric field autocorrelation function. ($g_1(\tau)$) [40]:

$$g_1(\tau) = g_1(0)e^{-q^2<\Delta r(\tau)^2>/6} \qquad (11.24)$$

where τ is the delay time, $g_1(0)$ is the autocorrelation function at τ=0, q is the scattering vector, q=4πn/λ sin (θ/2) and Δr(τ)² is the MSD of the tracer probe particle. Once the MSD is obtained, the viscoelastic properties such as the storage and viscous modulus $G'$, and $G''$ can be obtained from the Generalized Stokes Einstein Relationship (GSER) as highlighted in the earlier section.

The main difference between DLS based microrheology and DWS based microrheology is that the DLS technique is based on single scattering, whilst the DWS technique is based on multiple scattering. From an application perspective the DWS is more sensitive to small probe particle motion and allows access to higher frequencies then the DLS based microrheology technique. The concentration of tracer probe particles required is also different between the two techniques. As DWS works in the multiple scattering regime a high concentration of probe particles is required to ensure the right level of multiple scattering.

Particle Tracking Techniques-This is a relatively recent development in the microrheology area,[47] In this technique, the local viscoelasticity is obtained from the motion of the tracer particles. The MSD in this case is averaged over all particles and all video frames and is given by

$$< \Delta r^2(\tau) > = < \left(r(t+\tau) - r(t)\right)^2 >_{i,t} \qquad (11.25)$$

This carries out a quantification of the average distance moved by the probe particles in a time τ. In particle tracking microrheology, the MSD is obtained from the particle positions. The experimental set-up is very straight forward as it involves standard laboratory equipment such as an inverted fluorescence microscope and a CCD camera for image acquisition.

The MSD obtained through any of the above experimental techniques can itself provide insights into the underlying rheological properties of the complex fluids.

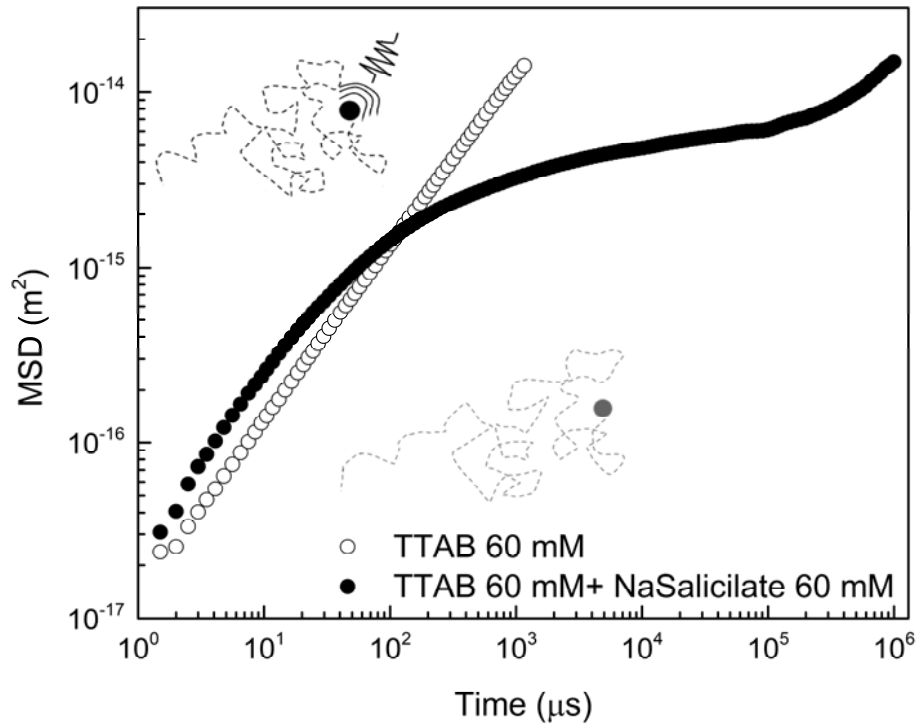

**FIGURE 11.6** Mean Squared Displacement (MSD) for a purely viscous fluid (red) exhibiting linear time dependence and for a viscoelastic fluid (blue) exhibiting deviation from the linear dependence. Cation tracer diameter = 220 nm.

The MSD obtained from a purely viscous complex fluid exhibits a linear dependence on time. As the complex fluid becomes more elastic, the particle motion becomes increasingly sub-diffusive and exhibits deviation from the linear behaviour. This is highlighted in figure 11.6.

The advances in microrheological techniques is continuing to evolve and additional techniques such as neutron spin echo, X-ray photon correlation spectroscopy, fluorescence correlation spectroscopy, Pulse Field Gradient NMR etc hold promise as potential future microrheological techniques.

### 11.4.4 Applications of Microrheology

Microrheology as highlighted in the introduction to this section has certain clear advantages over traditional mechanical rheometry. These include ability to gain insights into short time dynamics through access to high frequency data, ability to only utilize small sample volumes for carrying out measurements and gain insights into heterogeneities in dynamics and rheological behaviour. The following highlights some of the recent work which takes advantage of each of these benefits of microrheology

- ➢ Short Time Dynamics (High Frequency Data): Microrheology, especially DWS based optical microrheology allows access to viscoelasticity data over several decades in frequency, this includes access to data at very high frequencies such as $10^6$ Hz. These high frequencies allow access to the short time dynamics in complex fluids and is not accessible through mechanical rheometry. This insight is very useful for gaining insights into microstructural aspects in polymer solutions and self assembled surfactant systems such as wormlike micelles.[45,46,49] The high frequency data allows insights into the relaxation dynamics in these systems at short times which can be utilized in turn to extract out microstructural characteristics such as persistence length, contour length, entanglement length in wormlike micellar systems.[45,49]
- ➢ Small Sample Volume: Mechanical rheometers generally require large sample volumes especially when using geometries for low viscosity samples such as couette geometry. Even cone and plate requires significant sample volume. Additionally as these geometries are usually open to air, there can be significant evaporation, particularly when carrying out thermal studies. Both DLS and DWS microrheology is cuevette based requiring significantly lower sample volumes and because the sample is enclosed in a cuevette very little sample evaporation takes place. This is especially useful for biological samples such as proteins which are expensive and very little sample quantity is usually available to carry out characterization. These systems usually exhibit weak structural changes and rheometers do not have the sensitivity to pick up these small structural changes. As an example DLS microrheology has been applied to explore the onset of weak elasticity due to thermal aggregation of proteins.[40]
- ➢ Ageing and Heterogeneity in Rheological Behaviour: DLS and DWS based optical microrheology can be utilized to explore the glassy dynamics, aging in dense colloidal and structured colloidal systems such as Laponite.[48] The analysis of the MSD over time and different concentrations and compositions of Laponite for example gave insights into the changes in structure, dynamics and heterogeneity in these clay systems as a result of ageing.

### 11.4.5 Potential Issues in Microrheological Experiments

Microrheology can provide some unique insights as highlighted in the applications. The effective carrying out of robust and rigourous experiments however in microrheology is not straightforward. The main aspect that causes significant issues in microrheological measurements is the correct choice of probe particles. The key aspects that need to be taken into account include:[40]

- ➢ Particle Chemistry-the surface chemistry of the probe particle needs to be carefully considered such that there is no interaction between the particle and the complex fluid in which the particle is embedded. Any interaction would impact on the thermal motion of the probe particle and therefore on the extracted MSD.
- ➢ Particle size-the particle size is chosen such that it is larger then the relevant microstructural lengthscale of the complex fluid, such as the mesh size in a polymer.

This would allow the probe of the bulk rheological response. The particle size should however not be so large such that sedimentation becomes a problem. Sizes in the range of 0.5μm to 1.0 μm.
- ➢ Particle Concentration: The particle concentration needs to be chosen such that the scattering from the tracer particles dominates the scattering. In DWS microrheology the particle concentration chosen also needs to ensure that scattering is in the multiple scattering regime. The concentration needs to be balanced as too high concentration of the tracer particles can lead to the rheology of the complex fluid being impacted.

The experimental methodology to choose the particle characteristics for robust microrheological measurements is highlighted by Amin et.al.[40]

Langmuir. 25 (2009) 716–723. https://doi.org/10.1021/la802323x.